\documentclass[twoside,twocolumn,9pt]{article}
\usepackage{extsizes}
\usepackage[super,sort&compress,comma]{natbib} 
\usepackage[version=3]{mhchem}
\usepackage[left=1.5cm, right=1.5cm, top=1.785cm, bottom=2.0cm]{geometry}
\usepackage{balance}
\usepackage{mathptmx}
\usepackage{sectsty}
\usepackage{graphicx} 
\usepackage{lastpage}
\usepackage[format=plain,justification=justified,singlelinecheck=false,font={stretch=1.125,small,sf},labelfont=bf,labelsep=space]{caption}
\usepackage{float}
\usepackage{fancyhdr}
\usepackage{fnpos}
\usepackage[english]{babel}
\addto{\captionsenglish}{%
  
}
\usepackage{array}
\usepackage{droidsans}
\usepackage{charter}
\usepackage[T1]{fontenc}
\usepackage[usenames,dvipsnames]{xcolor}
\usepackage{setspace}
\usepackage[compact]{titlesec}
\usepackage{hyperref}

\usepackage{epstopdf}

\usepackage{amssymb}
\usepackage{siunitx}
\usepackage{xcolor}
\usepackage{siunitx}

\renewcommand{\vec}{\mathbf}
\renewcommand{\epsilon}{\varepsilon}

\definecolor{cream}{RGB}{222,217,201}

\begin{document}

\pagestyle{fancy}
\thispagestyle{plain}
\fancypagestyle{plain}{
\renewcommand{\headrulewidth}{0pt}
}

\makeFNbottom
\makeatletter
\renewcommand\LARGE{\@setfontsize\LARGE{15pt}{17}}
\renewcommand\Large{\@setfontsize\Large{12pt}{14}}
\renewcommand\large{\@setfontsize\large{10pt}{12}}
\renewcommand\footnotesize{\@setfontsize\footnotesize{7pt}{10}}
\makeatother

\renewcommand{\thefootnote}{\fnsymbol{footnote}}
\renewcommand\footnoterule{\vspace*{1pt}%
\color{cream}\hrule width 3.5in height 0.4pt \color{black}\vspace*{5pt}} 
\setcounter{secnumdepth}{5}

\makeatletter 
\renewcommand\@biblabel[1]{#1}            
\renewcommand\@makefntext[1]%
{\noindent\makebox[0pt][r]{\@thefnmark\,}#1}
\makeatother 
\renewcommand{\figurename}{\small{Fig.}~}
\sectionfont{\sffamily\Large}
\subsectionfont{\normalsize}
\subsubsectionfont{\bf}
\setstretch{1.125} 
\setlength{\skip\footins}{0.8cm}
\setlength{\footnotesep}{0.25cm}
\setlength{\jot}{10pt}
\titlespacing*{\section}{0pt}{4pt}{4pt}
\titlespacing*{\subsection}{0pt}{15pt}{1pt}

\fancyfoot{}
\fancyfoot[LO,RE]{\vspace{-7.1pt}\includegraphics[height=9pt]{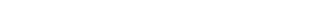}}
\fancyfoot[CO]{\vspace{-7.1pt}\hspace{13.2cm}\includegraphics{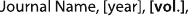}}
\fancyfoot[CE]{\vspace{-7.2pt}\hspace{-14.2cm}\includegraphics{head_foot/RF}}
\fancyfoot[RO]{\footnotesize{\sffamily{1--\pageref{LastPage} ~\textbar  \hspace{2pt}\thepage}}}
\fancyfoot[LE]{\footnotesize{\sffamily{\thepage~\textbar\hspace{3.45cm} 1--\pageref{LastPage}}}}
\fancyhead{}
\renewcommand{\headrulewidth}{0pt} 
\renewcommand{\footrulewidth}{0pt}
\setlength{\arrayrulewidth}{1pt}
\setlength{\columnsep}{6.5mm}
\setlength\bibsep{1pt}

\makeatletter 
\newlength{\figrulesep} 
\setlength{\figrulesep}{0.5\textfloatsep} 

\newcommand{\topfigrule}{\vspace*{-1pt}%
\noindent{\color{cream}\rule[-\figrulesep]{\columnwidth}{1.5pt}} }

\newcommand{\botfigrule}{\vspace*{-2pt}%
\noindent{\color{cream}\rule[\figrulesep]{\columnwidth}{1.5pt}} }

\newcommand{\dblfigrule}{\vspace*{-1pt}%
\noindent{\color{cream}\rule[-\figrulesep]{\textwidth}{1.5pt}} }

\makeatother

\twocolumn[
  \begin{@twocolumnfalse}
{\includegraphics[height=30pt]{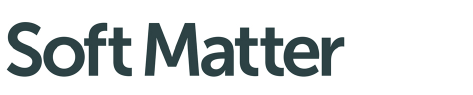}\hfill\raisebox{0pt}[0pt][0pt]{\includegraphics[height=55pt]{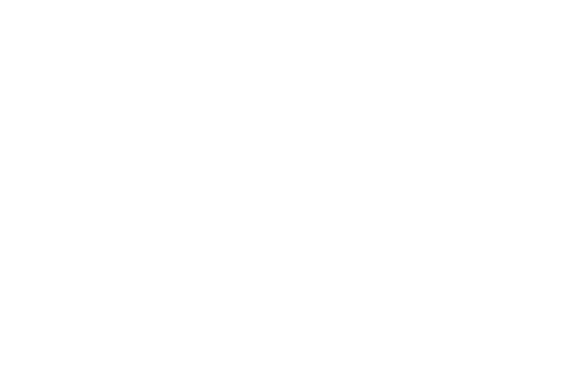}}\\[1ex]
\includegraphics[width=18.5cm]{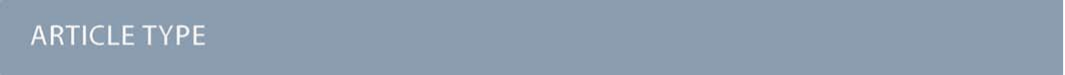}}\par
\vspace{1em}
\sffamily
\begin{tabular}{m{4.5cm} p{13.5cm} }

\includegraphics{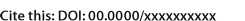} & \noindent\LARGE{\textbf{Probing microscopic dynamics in a uni-axially strained polymer network$^\dag$}} \\
\vspace{0.3cm} & \vspace{0.3cm} \\

& \noindent\large{N. H. P. Orr\textit{$^{a,b,\ddag}$}, G. Prevot \textit{$^{a}$}}, T. Phou \textit{$^{a}$}, and L. Cipelletti \textit{$^{a,c,\ast}$}\\

\includegraphics{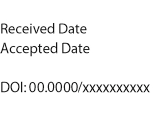} & \noindent\normalsize{We present a new apparatus that probes simultaneously the macroscopic mechanical response and the microscopic motion in polymer networks under uni-axial strain. The setup leverages photon correlation imaging, a space- and time-resolved dynamic light scattering method, to measure the dynamics along three orthogonal directions and on two distinct length scales, from tens of nanometers to a couple of microns. We show how to avoid artifacts due to scattering from the surface of the polymer films and  derive a theoretical expression for the intensity correlation function due to a purely affine deformation, showing that the setup sensitivity may be simply tuned by varying the acceptance angle of the collection optics. 
Finally, we demonstrate the capabilities of the setup by investigating the microscopic dynamics of a poly(dimethylsiloxne) polymer network under tensile strain in the linear regime. We find that non-affine dynamics dominate on length scales smaller than a few microns, above which the affine response is recovered. Surprisingly, the cross-over length separating the non-affine and affine regimes \textit{increases} upon decreasing the applied tensile strain. 
} \\

\end{tabular}

 \end{@twocolumnfalse} \vspace{0.6cm}

  ]

\renewcommand*\rmdefault{bch}\normalfont\upshape
\rmfamily
\section*{}
\vspace{-1cm}


\footnotetext{\textit{$^{a}$~Laboratoire Charles Coulomb (L2C), Université Montpellier, CNRS, Montpellier 34095, Montpellier, France. }}
\footnotetext{\textit{$^{b}$ CPCV, Department of Chemistry, École Normale Supérieure
PSL University, Sorbonne Université, CNRS
Paris 75005, France.}}
\footnotetext{\textit{$^{\ast}$}E-mail: luca.cipelletti@umontpellier.fr}

\footnotetext{\dag~Supplementary Information available: [details of any supplementary information available should be included here]. See DOI: 10.1039/cXsm00000x/}

\footnotetext{\ddag~ Present address: CPCV, Department of Chemistry
École Normale Supérieure
PSL University
Sorbonne Université
CNRS
Paris 75005, France }






\section{Introduction}
Network-forming soft matter systems are ubiquitous, from industrial products (tires, food, cosmetics…) to living organisms (e.g. in the cytoskeleton). Network-based materials often possess remarkable properties, such as high reversible deformability, light weightedness, and optical transparency. In particular, polymer networks comprise long chain molecules that may be extended, coiled up, physically entangled or chemically cross-linked to produce useful and varied mechanical properties, e.g. large elastic or plastic deformation capabilities. 

A thorough understanding of the behavior of soft materials under deformation is of utmost importance for directed material design and optimization of properties, within and beyond the linear regime where failure may eventually occur~\cite{barrat_soft_2024,cipelletti_microscopic_2020}. A full picture of soft solids under a mechanical drive requires simultaneous measurements at both microscopic and macroscopic length scales. In recent decades, microscopy coupled with rheology has been widely used to study a range of polymers and other network forming soft materials \cite{Villa2022, Edera2017, Edera2021, Basu2011, Wen2012, Hepworth2001}. In tandem with recent advances in computational and analytical modeling \cite{Heussinger2007,Tauber2021,Chen2023,Panyukov2023, Assadi2025,walker_toughness_2025} a more thorough understanding of the interplay between microscopic and macroscopic behavior is emerging. For example, recently simulations have shown how load sharing between the two
networks within a double network elastomer leads to the delocalization of stress so that a double network inherits both the stiffness of its brittle first network and the ductility of its soft second network hindering cascades of correlated bond breakage \cite{Tauber2021, walker_toughness_2025}.

Scattering techniques lack the sub-micron spatial resolution of microscopy, but they enable a wider range of length scales and larger sample volumes to be probed. Furthermore, unlike microscopy, they do not require the material constituents to be of size comparable to the wavelength of visible light, allowing for measurements on a much wider range of systems. While conventional dynamic light scattering (DLS) in the far field averages over the whole volume illuminated by a laser beam and over the experiment time, more recent, powerful approaches such as differential dynamic  microscopy \cite{Cerbino2008} and photon correlation imaging \cite{duri_resolving_2009} allow for time- and spatially-resolved experiments. As such, they are particularly insightful when materials deform heterogeneously, e.g. in the non-linear stress versus strain regime near and during material failure. 

Much of the previous work has concentrated on the multiple scattering regime probed by diffusing wave spectroscopy, DWS~\cite{weitz_diffusing-wave_1993}. A remarkable feature of DWS is its extreme sensitivity to small-scale motion, down to a fraction of nanometers, which enables very small deformation fields to be accurately quantified~\cite{wu1990,erpelding2008,erpelding2013,Nagazi2017,vanderkooij2018,Ju2021}. Single scattering experiments, however, remain particularly attractive for several reasons. First, many polymer materials are almost transparent, so that they may be directly probed by DLS, while DWS typically requires adding significant amounts of tracer scatterers, usually colloidal particles embedded in the material, which may alter its mechanical properties. Tracer particles might still be added to improve the DLS signal for very transparent materials, but in significantly smaller amounts as compared to DWS. Second, single scattering can discriminate between the affine and non-affine contributions to microscopic motion~\cite{Aime2019,Aime2019II,milani_space-resolved_2024}, a highly desirable feature since the affine displacement is of little interest and can be directly obtained from the macroscopic sample deformation. This is not the case in DWS: as photons undergo many scattering events, they travel through the material in a random walk way, thereby averaging microscopic displacements in all directions, including motion due to the affine deformation field~\cite{wu1990,erpelding2008,ju_multispeckle_2022}. Finally, single scattering is the relevant regime for X-photon correlation spectroscopy (XPCS), the equivalent of DLS using synchrotron coherent X-rays, where experiments probing microscopic dynamics simultaneously to mechanical measurements are increasingly popular, see e.g. Ref.~\cite{leheny_rheo-xpcs_2015}.

Notably, only shear deformations have been thoroughly examined 
in single scattering experiments~\cite{ali2016, Edera2017, Aime2018, Aime2019, Aime2019II, Larobina2021, Edera2021, Villa2022}, with the exception of the XPCS investigation of the dynamics of filler particles in an elastomer~\cite{ehrburger-dolle_xpcs_2012}. However, for many materials tensile tests are more relevant. We fill this gap by presenting here a new setup for simultaneously measuring the mechanical properties and microscopic dynamics of self-standing polymer films in tensile tests. The microscopic dynamics are quantified by photon correlation imaging, PCI \cite{duri_resolving_2009}, a dynamic light scattering method in the single scattering regime that leverages CMOS cameras to provide spatial maps of the dynamics between arbitrary pairs of experimental times $t$ and $t+\tau$. We discuss possible artifacts arising from surface scattering and how to mitigate them. We demonstrate the capabilities of the new setup with measurements of the microscopic dynamics in poly(dimethylsiloxane), PDMS, networks during uni-axial extension tests at constant pulling speed, unveiling an unexpected strain-dependent contribution of non-affine dynamics even at the smallest applied strains.

The rest of the paper is organized as follows: in Sec.~\ref{sec:setup} we detail the PCI-based setup, summarize our analysis routines and describe the PDMS polymer sample preparation. In Sec.~\ref{sec:theory}, we derive the theoretical form of the two-time intensity correlation function $g_2-1$ measured by PCI for purely affine motion, briefly mentioning how non-affine motion may accelerate the decay of $g_2-1$. We present our experimental results in Sec.~\ref{sec:results}, starting by discussing the mitigation of surface scattering in order to probe microscopic motion from the bulk of the sample (Secs.~\ref{sec:surface} and~\ref{sec:bulk}). In Sec.\ref{sec:pre-strain} we discuss how microscopic motion depends on the probed  direction and length scale, unveiling an unexpected dependence on the applied macroscopic pre-strain. Finally, in Sec.\ref{sec:conclusion} we briefly discuss and summarize our findings.

\section{Experimental setup}
\label{sec:setup}

\subsection{Instrumentation}

\begin{figure}[H]
\centering
\includegraphics[width=\linewidth]{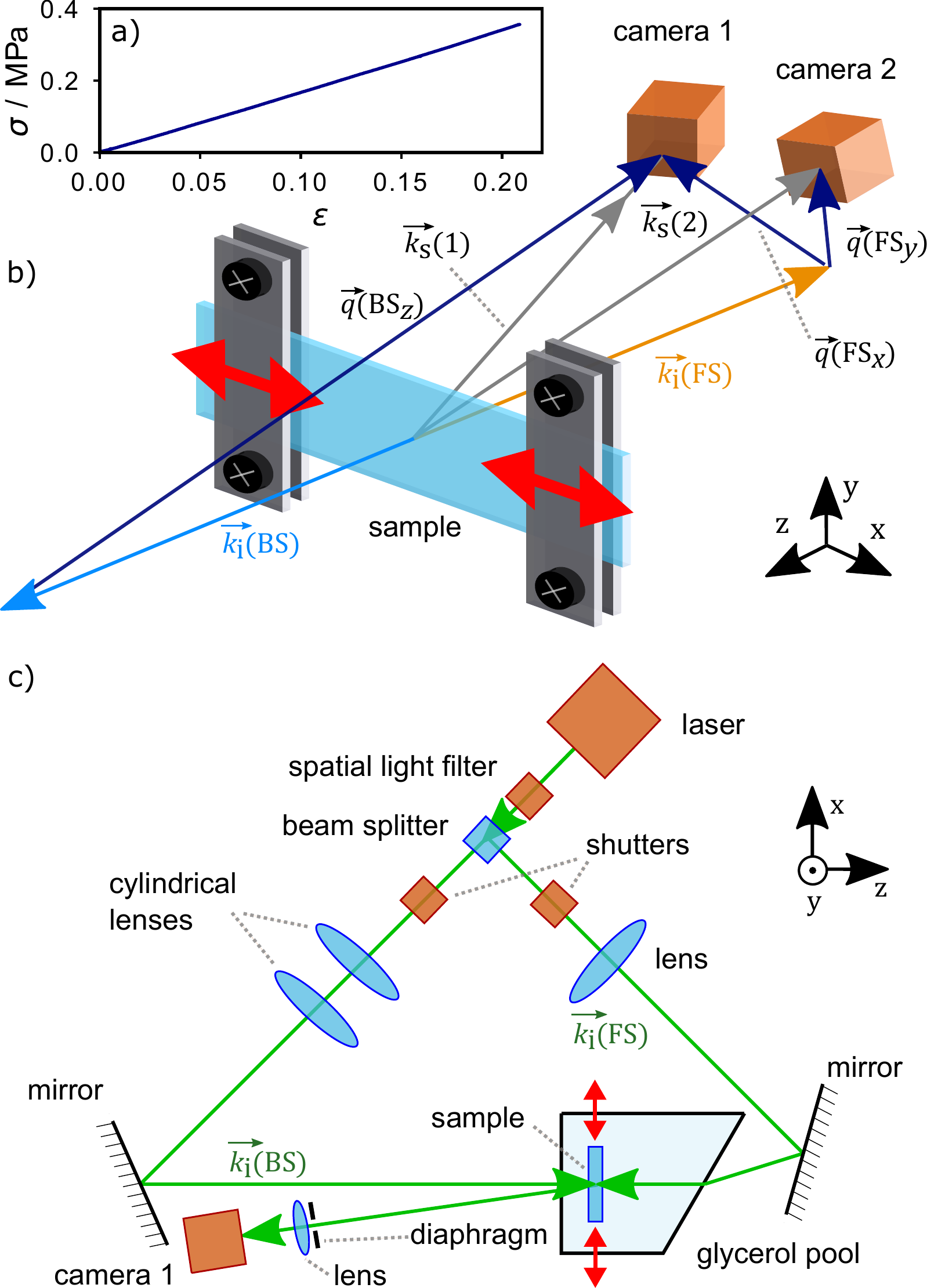}
\caption{a) True stress $\sigma$ as a function of true strain $\varepsilon$, for a typical PDMS sample seeded with $3 \times 10 ^{-4}$ mass fraction of melamine resin colloidal particles. b) A schematic view of the of the incoming ($\vec{k}_i$) and scattered ($\vec{k}_s$) wave vectors, and of the scattering vectors ($\vec{q}$), for the forward (FS) and back-scattered (BS) geometries described in the text. The red arrows show the strain direction. Note that, for clarity, the scattering angles are not accurate and that the second $\vec{q}$(BS$_z$) vector has been omitted. 
c) Top view of the optical paths for BS and FS. The position of camera 2 and the related detection optical paths have been omitted for clarity. They sit directly above the beam path $\vec{k}_{\mathrm{i}}($BS$_z)$ just next to camera 1.}
    \label{fig:setup}
\end{figure}

Our custom apparatus is shown schematically in Fig.~\ref{fig:setup}b and c. It consists of a mechanical part, comprising two motors, a displacement detector and a force sensor, allowing for stress and strain measurements, and an optical part, for space- and time-resolved dynamic light scattering (PCI).  
We resolve motion on different length scales and along different directions by collecting light at three different scattering vectors (dark blue arrows in Fig.~\ref{fig:setup}b). To this end, laser light at 532 nm (HÜBNER Photonics Cobalt Samba, 100 mW) first passes through a spatial filter containing a parabolic lens (diameter $ d= 3~\mathrm{mm}$, focal length $f = 2~\mathrm{mm}$, Thorlabs C151TMD-A) and a tungsten pinhole with $d = 5~\mu\mathrm{m}$. The filtered laser beam is split into two paths, for forward, FS, and backward, BS, scattering, respectively. A $10~\mathrm{mm}$ wide 10:90 (R:T) non-polarizing beam splitter cube (Thorlabs BS037) directs the diverging beam to the FS and BS paths, respectively, see Fig.~\ref{fig:setup}c. In the FS path, the beam is collimated by an $f = 80 ~\mathrm{mm}$ 
plano-convex lens, producing an expanded beam with $1/e^2$ diameter of $\approx 50 ~\mathrm{mm}$. 
Similarly, in the BS channel the beam is collimated by two crossed cylindrical lenses, with focal lengths of $f = 100$ and $f = 300 ~\mathrm{mm}$, respectively. The shape of the thus collimated beam is oval with major and minor axes with $1/e^2$ lengths of $\approx 20 ~\mathrm{mm}$ and $\approx 60 ~\mathrm{mm}$, respectively. Note that the intensity of back-scattered light is typically lower than that for forward scattering: the choice of lenses and of the 10:90 intensity beam splitter was made so that the cameras can use the same exposure time for light coming from the BS and FS channels. Finally, the collimated beam is directed towards the sample by broadband, high reflectance mirrors (Newport 20Q620BB.HR2), with $d=50.8~\mathrm{mm}$, see Fig.~\ref{fig:setup}c.

The scattered light is collected along two paths. Each passes a Newport diaphragm placed in front of an $f=80~\mathrm{mm}$ plano-convex objective lens. These lenses form a speckled image of the illuminated sample onto the ams CMV2000 CMOS sensors of two Basler acA2000-340km Camera Link cameras, with 5 times magnification for both FS and BS. 
The purpose of the diaphragms is to control the acceptance angle of the collection optics, thereby setting the speckle size~\cite{goodman_speckle_2007-1}, which is typically comparable to the pixel size. The sample image is formed by light scattered with different scattering vectors $\vec{q}$, depending on the illuminating and collection paths. Three independent scattering vectors are probed: $\vec{q}(\mathrm{FS}_x)$, $\vec{q}(\mathrm{FS}_y)$ and $\vec{q}(\mathrm{BS}_z)$, which are approximately parallel to their respective Cartesian directions, $x, y$ or $z$, Fig.~\ref{fig:setup}b.
The magnitude of the two FS vectors is approximately equal to each other with a value of $q_{\mathrm{FS}} = \qty{2.0}{\micro \metre} ^{-1}$ resulting in a sensitivity of microscopic displacements within the polymer network of the order of $\pi/q_{\mathrm{FS}} = \qty{1.6}{\micro \metre}$. Likewise, the magnitude of the BS vectors is $q_{\mathrm{BS}} = \qty{34}{\micro \metre} ^{-1}$, with a corresponding probed length scale of approximately $90~\mathrm{nm}$, 17.8 times shorter than for FS. Note that, due to the large scattering angle and the small distance between cameras, the $\vec{q}$ vectors of both BS cameras are approximately equal and will be treated as such herein. In practice, all BS data were collected with camera 1, except for those in Figs.~\ref{fig:surface}c,\ref{fig:pre-strain}e and g. The geometry of the scattering vectors with respect to the sample is shown in Fig.~\ref{fig:setup}b. The components of $\vec{q}$ are $[q_x,q_y,q_z] = [2.1,~1.8 \times 10^{-3},~0.13]~\qty{}{\micro\meter^{-1}}$ for FS$_x$, $[q_x,q_y,q_z] = [4.8 \times 10^{-4},~1.9, ~6.8 \times 10 ^{-2}]~\qty{}{\micro\meter^{-1}}$ for FS$_y$ and $[q_x,q_y,q_z] = [1.5,~8.1\times 10 ^{-2},~34]~\qty{}{\micro\meter^{-1}}$ for BS$_z$ (camera 1), and $[q_x,q_y,q_z] = [0.32,~1.5,~34]~\qty{}{\micro\meter^{-1}}$ for BS$_z$ (camera 2), where the sample lays in the $xy$ plane and $x$ is the pulling direction. 

Image acquisition is done by alternating between FS and BS illumination, using two shutters, one of which is custom built, while the other is a Newport electronic shutter (76992) with a $6~\mathrm{mm}$ aperture. 
The shutters are placed directly after the beam splitter along each path and are synchronized with the cameras acquiring speckle images. The shutter triggering and camera acquisition are controlled via custom software running on a PC connected to the setup.

To explore the role of speckle size on the decay of the correlation function, discussed in Sec.~\ref{sec:bulk}, we use a modified setup that allows for acquiring images with different speckle sizes at the same scattering angle. The setup is schematically shown in Fig.~\ref{fig:speckle_size}c: forward scattered light passes through a 50:50 non-polarizing beam splitter cube, 2 inches in width (Thorlabs BS031), before being collected by objective lenses on each branch of the optical path. In this case, the diaphragm in front of the objective lenses was set to a different aperture to obtain images with different speckle size on each camera. 

Simultaneously to PCI, we perform uni-axial tensile tests, thanks to a custom-built, strain-controlled rheometer. Our PDMS samples (dimensions in $x$, $y$, and $z$ of $ \approx 25~\mathrm{mm}, 10~\mathrm{mm}$ and between $0.32~\mathrm{mm}$ and $2.85~\mathrm{mm}$, respectively) are held by two clamps. Strain is applied by displacing the clamps along the $x$ direction, using translation stages (Newport M-UMR5.25), set in motion by two  motorized actuators (Newport LTA-HL). Unless stated otherwise, the two motors pull in opposite directions at speeds of $1.25~\mu\mathrm{ms}^{-1}$ and $5~\mathrm{nms}^{-1}$ for FS and BS experiments, respectively. This results in very small imposed strain rates of $\approx 1\times 10^{-4}~\mathrm{s}^{-1}$ and $\approx 4\times 10^{-7}~\mathrm{s}^{-1}$ for FS and BS, respectively. The actual clamp displacement is recorded by a position detector (Keyence IL-065 laser head with an IL-1000 sensor). Throughout this paper, we quantify deformation using the true strain $\epsilon(t) = \ln[1 + \Delta L(t) /L_0]$ where $\Delta L(t) = L(t) - L_0$ and $L_0$ are the change in and the macroscopic length, respectively, in the $x$ direction. An Andilog Centor Star Touch force gauge and SBlock force load cell measure the resistance force exerted by the sample. True stress is calculated as $\sigma = F / (A-\nu \varepsilon)$, where $F$ is the magnitude of the force measured from the meter, $A$ is the initial cross-sectional area in the $yz$ plane, and $\nu$ is the Poisson ratio, describing the response along directions orthogonal to the applied strain. For our PDMS samples, we measure $\nu = -\frac{\Delta y / y_0}{\Delta x / x_0} = 0.42$ by image analysis of small deformations in the $xy$ (sample) plane. We assume that $\nu$ is the same in the $yz$ plane.
As shown by the stress-strain curve of Fig.~\ref{fig:setup}a, in the range of deformations investigated here the sample behaves linearly, with a Young's modulus of $1.8~\mathrm{MPa}$.

As discussed in detail in Sec.~\ref{sec:surface}, single scattering experiments require enhancing the scattering from the bulk of the sample as compared to that from its surface. To this end, we incorporate a small amount of melamin resin colloidal particles to the polymer network (MicroParticles GmbH, diameter 418 nm, mass fraction $3 \times 10^{-4}$). Furthermore, we immerse the clamps and sample in a custom-built pool filled with glycerol, which reduces the refractive index mismatch between the sample and the surrounding medium, thereby drastically reducing surface scattering. The pool is shaped to direct reflections of the illuminating beam away from the cameras, see Fig.~\ref{fig:setup}c. We checked that the mechanical properties of the polymer films measured by our setup are not significantly affected neither by the incorporation of particles$^{\dag}$ nor by the presence of glycerol, as expected due to the trace amount of particles and the very slow pulling speed. 

\subsection{Photon correlation imaging analysis}
In PCI, a sample image is formed using light scattered in a narrow cone centered around a direction corresponding to a well-defined scattering vector $\vec{q}$. Due to the use of coherent laser light and the small acceptance angle of the imaging optics, the images have a speckled appearance, see Fig.~\ref{fig:speckle_size}c. Each speckle corresponds to a small, approximately cylindrical scattering volume with a length spanning the thickness of the sample. The cylinder cross-section can be increased (decreased) by closing (opening) the diaphragm in front of the objective lenses, see Fig.~\ref{fig:setup}c. Microscopic motion results in the speckle intensity fluctating in time, as in conventional DLS~\cite{Berne2000}. These fluctuations are quantified by a two-time intensity correlation function, calculated for each speckle (or camera pixel) and averaged over several pixels~\cite{Duri2005,duri_resolving_2009}, a well-established method referred to as multispeckle DLS~\cite{kirsch_multispeckle_1996}. Normally, this correlation function is a function of lag time, $\tau$, and time, $t$. However, in our uni-axial strain experiment the dynamics are fully dominated by those induced by the imposed strain. Therefore, we express the intensity correlation function as a function of strain and strain increment: 
\begin{equation}
    g_2(\vec{q}, \varepsilon, \Delta \varepsilon) - 1 = \beta \dfrac{\langle I_p(\varepsilon)I_p(\varepsilon + \Delta \varepsilon) \rangle_{\vec{r}}}{\langle I_p(\varepsilon) \rangle _{\vec{r}} \langle I_p(\varepsilon + \Delta \varepsilon) \rangle_{\vec{r}}} - 1\,,
\label{eq:g2-1}
\end{equation}
where $\varepsilon$ and $\varepsilon + \Delta \varepsilon $ are the true strain at times $t$ and $t + \Delta t$, respectively. 
$I_p$ is the intensity of a given pixel and $<\cdot \cdot \cdot>_{\vec{r}}$ is an average over all pixels in a small region of interest (ROI) centered around position $\vec{r}$ in the sample. 
Typically, we use ROIs of size 53 $\times$ 103 pixels ($x \times y$), corresponding to $1.33 \times 2.58~\qty{}{\milli\meter^{2}}$ on the sample. The prefactor $\beta$ insures that $g_2-1 \rightarrow 1$ for $\Delta \epsilon \rightarrow 0$. To reduce the statistical noise due to the finite number of speckles in the ROI, $\beta$ is slightly modulated for each pair of images that are correlated, as explained in Ref.~\cite{Martinelli2024}. As for conventional DLS, the intensity correlation function $g_2-1$ is the square of the intermediate scattering function that quantifies microscopic dynamics~\cite{Berne2000,lindner_interacting_2024}.

\subsection{Materials}
\label{sec:materials}
Pure PDMS samples were prepared with a 9:1 mass ratio of polymer (SYLGARD 184 - BASE) to curing agent, degassed under vacuum and then poured into polystyrene petri-dishes with an inverted lid using spacers to produce sheets of varying thicknesses. They were then cured in a $60^\circ$C oven for 2 hours. 
PDMS containing $3\times 10^{-4}$ mass fraction melamine formaldehyde colloidal particles were prepared by dispersing melamine resin particles (Microparticles GMBH) with diameter $418~\mathrm{\mu m}$ first in acetone using sonication and magnetic stirring. Then, the acetone and particle dispersion was added to the PDMS polymer and mechanically stirred at $10000$~rmp for 20 minutes. The mixture was placed under vacuum until the acetone evaporated. Curing agent, again with a 9:1 ratio, was added and mixed well and degassed under vacuum. Sheets of various thicknesses were prepared as before. Particles are added to the samples to increase bulk scattering as discussed in Sec.~\ref{sec:surface}. The particle size is much larger than the mesh size of PDMS, $10 - 16~\mathrm{nm}$, as reported using AFM \cite{Drebezghova2020} on similarly crosslinked samples. One can also estimate the mesh size using entropic rubber-elastic scaling where the mesh size $\xi \approx \left(\frac{K_\mathrm{B}T}{E}\right)^{\frac{1}{3}}$. With a Young's modulus, $E = 1.8~\mathrm{MPa}$, we obtain a smaller mesh size $\xi \approx 1~\mathrm{nm}$, again much smaller than the added microparticle size. Therefore, we expect the motion of the microparticles to be representative of the polymer network.

\section{Theory: the intensity correlation function under uni-axial strain}
\label{sec:theory}
The intensity correlation function $g_2-1$ may decay due to various sources of microscopic and mesoscopic motion: spontaneous dynamics, affine and non-affine deformation due to the uni-axial deformation, and the drift of any given ROI while extending the sample. The spontaneous dynamics, i.e. the dynamics that would be measured for a sample at rest, is negligible on the time scale of our experiments. Indeed, thermal fluctuations are very restrained because of the high Young's modulus and because no network rearrangements can be thermally activated, since the polymers are cross-linked by covalent bonds. The contribution owing to the local drift upon stretching is due to the fact that in PCI any rigid motion of the sample is mirrored by the motion of its image on the detector. It can be corrected for, either by using image registration techniques~\cite{cipelletti_simultaneous_2013}, or by analyzing a ROI in the center of the sample, which is stationary in the laboratory reference frame. We shall thus focus on the contribution due to purely affine motion under uni-axial extension, which we expect to be dominant in the linear regime probed here, before considering how additional non-affine motion may affect the measured correlation functions.

\subsection{A purely elastic material: contribution of the affine displacement field}
In the classical theory of rubber elasticity \cite{ Kuhn1936, Wall1942, Flory1943, Rubinstein2003}, deformation is assumed to be affine, where motion is distributed spatially homogeneously across all length scales. In this section we derive the form of the intensity correlation function $g_2 - 1$ for affine motion under uni-axial strain.  

For single scattering, the intensity correlation function may be written in terms of the sum over the contributions of all scatterers, or, equivalently, as the Fourier transform of the probability distribution $P(\Delta \vec{r})$ of the scatterers' displacement~\cite{Berne2000,pusey_dynamic_2024}. For strain-induced dynamics, one has
\begin{equation}
        g_{2}(\vec{q}, \Delta\varepsilon) - 1 = 	\left | \int_{V_s} P[\Delta \vec{r}(\Delta \varepsilon)] e^{i \vec{q} \cdot \Delta \vec{r}(\Delta \varepsilon)} \mathrm{d}^3 \Delta \vec{r} 	\right |^2 \,,
        \label{eq:integralvolumeform}
\end{equation}
where we have omitted the dependency on pre-strain $\epsilon$ for simplicity, and where $\Delta \varepsilon$ is the strain increment and the integral is over the scattering volume $V_s$. Since $g_2-1$ is first calculated for a given pixel (see Eq.~\ref{eq:g2-1}), i.e. for a given speckle, and then averaged over a ROI, $V_s$ is to be identified with the sample volume associated to a speckle. As mentioned earlier, this may be approximated by a cylinder of height $l_z$ equal to the sample thickness, of the order of 1-2 mm, and of later dimension $l_s$, of the order of several tens of $\qty{}{\micro \metre}$. In writing Eq.~\ref{eq:integralvolumeform}, we made two assumptions: first, the scattering intensity is the same for all scatterers; second $g_2-1$ may be safely approximated by its self (or incoherent) part. Both conditions are fulfilled for our samples, where scattering is dominated by the signal from the identical, randomly dispersed colloidal trace particles. Usually, DLS probes collective motion and $g_2-1$ is related to the coherent intermediate scattering function. However, the coherent and incoherent scattering functions become equivalent for randomly distributed scatterers or when the magnitude of the scattering vector is larger than $1/D$, where $D$ is the average distance between scatterers\cite{Berne2000}. In our case, both conditions are fulfilled, as colloidal particles are incorporated into the network in random positions and $ {q}_{\mathrm{BS}} > {q}_{\mathrm{FS}} > \frac{1}{D} = \qty{0.32}{\micro \meter}^{-1} $.

For affine motion, the probability distribution of displacements factors out in its components along $x$, $y$, and $z$: $P(\Delta \vec{r}) = P_x(\Delta x)P_y(\Delta y)P_z(\Delta z)$. We consider the displacement of a scatterer initially at position $(x,y,z)$. As shown in the Supplementary Information$^{\dag}$, upon a macroscopic true strain increment $\Delta \epsilon$ along $x$ and assuming a perfectly elastic sample with Poisson's ratio $\nu$, the scatterer displacement is
\begin{equation}
    \begin{split}
        \Delta x &= x\Delta\varepsilon \\
        \Delta y &=  -\nu y \Delta\varepsilon \\
        \Delta z &=  -\nu z \Delta\varepsilon  \,.
    \end{split}
\label{eq:delta_xyz}
\end{equation}
Here, we have assumed that the strain increment is sufficiently small for the increment of true strain $\Delta \epsilon$ to be a good approximation of the incremental engineering strain $\Delta L/L(t)$, see the Supplementary Information$^{\dag}$ for details. This approximation is well justified, since we shall see that typically $g_2-1$ fully decays over strain increments $\Delta \epsilon << 1$.

In order to determine simple expressions for $P_x$, $P_y$, $P_z$, we approximate $V_s$ by a square prism of side $l_s$ and height $l_z$. Since the probability distribution for the scatterers' position is constant within $V_s$, using Eqs.~\ref{eq:delta_xyz} one has
\begin{equation}
    \begin{split}
        P_x&=\dfrac{1}{\Delta \varepsilon l_s}  \quad -\frac{\Delta \varepsilon l_s}{2} < \Delta x < \frac{\Delta \varepsilon l_s}{2}  \\ 
        P_y&=\dfrac{1}{\Delta \varepsilon \nu l_s} \quad -\frac{\Delta \varepsilon \nu l_s }{2} < \Delta y < \frac{\Delta \varepsilon \nu l_s }{2} \quad \\
        P_z&=\dfrac{1}{ \Delta \varepsilon \nu l_z} \quad -\frac{\Delta \varepsilon \nu l_z }{2} < \Delta z < \frac{\Delta \varepsilon \nu l_z }{2} \,.
    \end{split}
\label{eq:PxPyPz}
\end{equation}
Note that in Eqs.~\ref{eq:PxPyPz} we have assumed that the average displacement is zero: as discussed previously, this condition can be fulfilled either by correcting for any drift motion the PCI images~\cite{cipelletti_simultaneous_2013}, or by considering a ROI in the center of the sample. By inserting Eqs.~\ref{eq:PxPyPz} in Eq.~\ref{eq:integralvolumeform} and
using the identity $a^{-1}\int_{-a/2}^{a/2} \mathrm{e}^{iqx} dx= \mathrm{sinc}(aq/2)$, one obtains
\begin{equation}
        g_2(\vec{q}, \Delta\varepsilon) - 1 = 	\bigg | \mathrm{sinc} \bigg(\frac{\Delta \varepsilon l_s q_x}{2} \bigg) \mathrm{sinc} \bigg(\frac{\Delta \varepsilon l_s  \nu q_y}{2} \bigg) \mathrm{sinc} \bigg(\frac{\Delta \varepsilon l_z \nu q_z}{2} \bigg)	\bigg |^2 \,,
\label{eq:bulk_scatter}
\end{equation}
where $\mathrm{sinc}(x) = \sin(x)/x$.

Importantly, Eq.~\ref{eq:bulk_scatter} shows that not only does the correlation function depend on $\Delta \varepsilon$ and the components of the scattering vector, but also on $l_s$, $l_z$ and $\nu$. Therefore, careful attention is required to probe rearrangements along a desired direction within the sample. For example, to probe rearrangements parallel to the direction of extension, $x$, an experimentalist must verify that $l_s q_x > \nu  l_z q_z$, which for sufficiently thick samples ($l_z \approx 2$~mm) and small speckle sizes ($l_s \approx \qty{50}{\micro \meter}$), may not be the case. 
These considerations will be made in Figs.~\ref{fig:speckle_size}b and Fig.~\ref{fig:pre-strain}b,d,f, where the three sinc$^2$ terms of Eq.~\ref{eq:bulk_scatter} will be plotted separately, as a reference against which experimental data should be compared.

\subsection{Contribution of non-affine displacements}

Affine deformation ignores any spatial correlations and therefore couplings between the local network structure and the mechanical properties\cite{Heussinger2007}. However, non-affine displacements may stem from spatial fluctuations of the elastic modulus~\cite{Basu2011} or from additional local degrees of freedom, such as polymer chain displacements that are activated by the strain, even in the linear elastic regime.  
Furthermore, beyond the linear regime, non-affinity may arise from local rearrangements in a damaged zone and from the elastic response to a local damage of the rest of the network. 

Quite generally, non-affine displacements speed up the decay of $g_2-1$, because they introduce additional sources of dynamics. Therefore, Eq.~\ref{eq:bulk_scatter}, the theoretical expression for a purely affine deformation, represents the limiting case of the slowest decay that may be observed for a strained material. The actual decay observed in experiments is typically faster~\cite{milani_space-resolved_2024}, as  we shall show in Secs.~\ref{sec:bulk} and \ref{sec:pre-strain} for our PDMS samples.  Decoupling the contributions of affine and non-affine dynamics is probably one of the greatest challenges in single scattering measurements on driven systems~\cite{Aime2018,Aime2019,Pommella2019,chen_microscopic_2020,edera_deformation_2021,milani_space-resolved_2024}. Data analysis may be greately simplified by assuming that non-affine and affine displacements as uncorrelated, as hypothesized in previous works~\cite{Basu2011,Aime2018,Aime2019,edera_deformation_2021,milani_space-resolved_2024}, or if the decay times of the affine and non-affine contributions are well separated~\cite{fuller_measurement_1980}. Under these assumptions, the intensity correlation function factors in the product of two terms, one accounting for affine and the other for non-affine contributions, respectively. Even if this assumption may not hold, the ratio of the experimental $g_2-1$ to the theoretical form of Eq.~\ref{eq:bulk_scatter} provides a convenient means to  asses the importance of non-affine dynamics.

\section{Results}
\label{sec:results}
\subsection{Surface scattering causes spurious oscillations in the correlation function}
\label{sec:surface}
\begin{figure}[h]
    \centering
    \includegraphics[width=\linewidth]{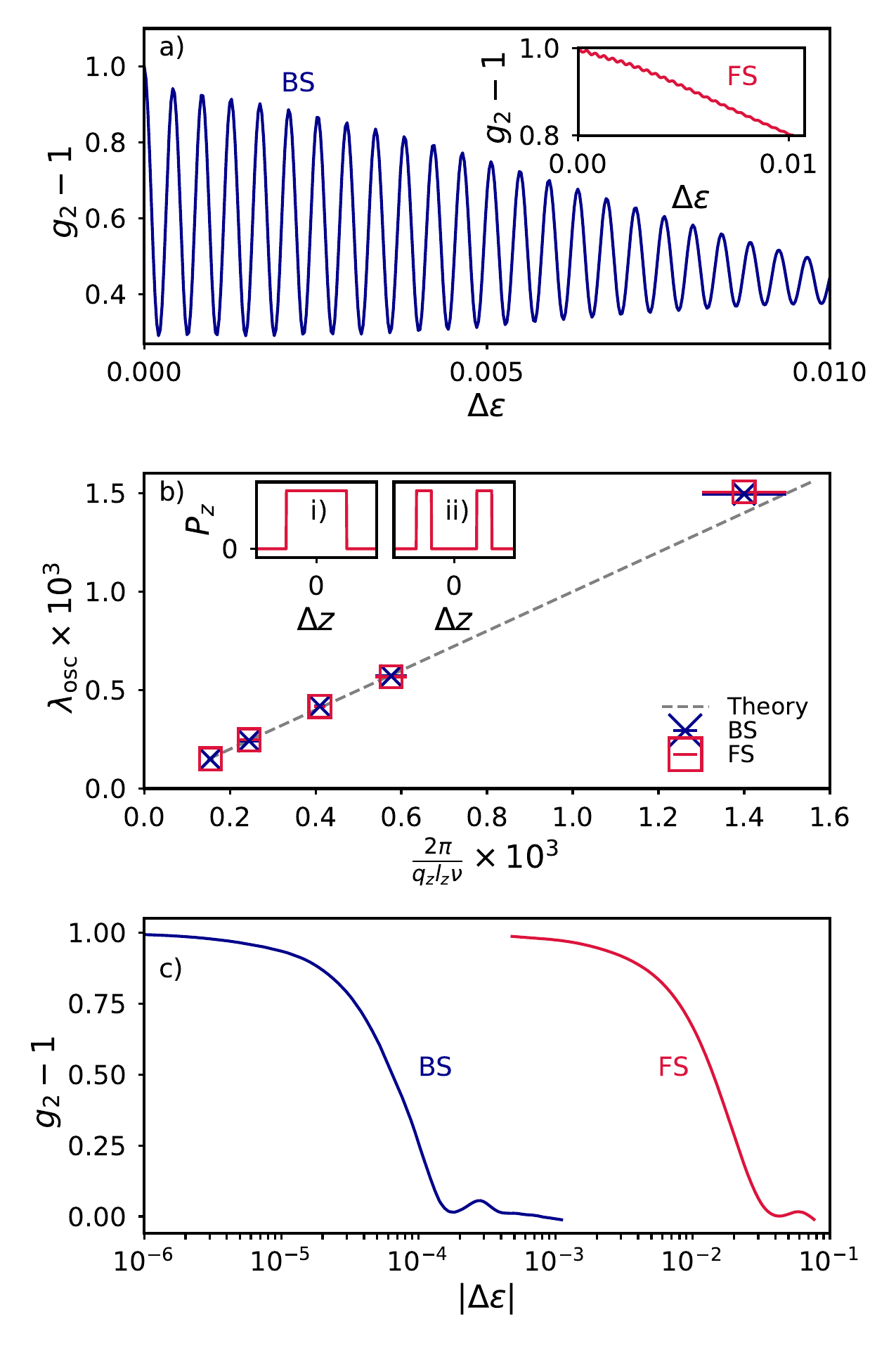}
    \caption{a) Main plot: $g_2-1$ for an uni-axially strained pure PDMS sample in air measured in the BS configuration. Inset: same for FS. b) Main plot: Predicted (dashed line) and measured (symbols) non-dimensional wavelength of the oscillations of $g_2-1$, for samples of various thicknesses, $l_z$. The error bar is the standard error. Inset: schematic shape of the scattered intensity-weighted probabilities of the displacements along $z$ for i) the bulk scattering case and ii) the surface scattering case. 
    In both a) and b) the two motors were pulling the sample at a speed of $ 1.5~\mu\mathrm{ms}^{-1}$ in opposite directions, imposing a strain rate of $\approx 7\times 10^{-5}~\mathrm{s}^{-1}$. c) Intensity correlation functions measured when surface scattering is strongly reduced and bulk scattering is increased, as detailed in the text. Data for a $1.77~\mathrm{mm}$ thick PDMS sample. }
    \label{fig:surface}
\end{figure}

We start by discussing $g_2-1$ measured for a pure PDMS sample in air under uni-axial strain. The resulting BS correlation function, shown in Fig.~\ref{fig:surface}a, has a surprising form. Instead of decaying following the $\mathrm{sinc}^2(x)$-like shape predicted by Eq.~\ref{eq:bulk_scatter}, it exhibits wide oscillations. Similar oscillations, of smaller amplitude but with the same frequency, are also seen in the FS correlation function, inset of Fig.~\ref{fig:surface}a.

Tests with samples for which surface scattering was deliberately enhanced by roughening their surface indicate that these oscillations result from scattering from the sample surfaces. As the sample is stretched, the two surfaces parallel to the $xy$ plane approach each other, due to the positive Poisson's ratio, causing interferences that oscillate from constructive to destructive as the sample is increasingly strained.
This phenomenon is reminiscent of the oscillations of the transfer function of a Fabry-Perrot interferometer~\cite{born_principles_2013}, with the two surfaces of the polymer film acting here similarly to the reflecting surfaces in a Fabry-Perrot cavity. However, we emphasize that here the phenomenon is caused by scattering, not by the reflection of the incident beam, since the cameras are positioned away from specular reflection. Rather, light scattered by the first surface encountered by the incident beam interferes with light scattered by the second surface, which is illuminated by the light transmitted through the almost transparent sample. Note that surface scattering is likely to occur, because the polymer samples are not optically flat: any imperfections in the mold used for sample preparation will be imprinted on its surfaces.

To rationalize these oscillations, we derived a modified version of Eqs.~\ref{eq:PxPyPz} and~\ref{eq:bulk_scatter} assuming that scattering from the bulk is negligible as compared to that originating from a thin layer of thickness $l_{\mathrm{surf}} << l_z$ at both sample surfaces. Accordingly, $P_z$ has to be modified as sketched in the inset of Fig.~\ref{fig:surface}b. Mathematically, the modified $P_z$ may be expressed as
\begin{equation}
    \begin{split}
        P_z &= G_z \circledast H_z \\
        G_z &= \dfrac{1}{l_\mathrm{surf}} , \quad -\dfrac{l_\mathrm{surf}}{2} < x < \dfrac{l_\mathrm{surf}}{2} \\
        H_z &= \dfrac{1}{2} \left[ \delta \Bigr ( \Delta z - \dfrac{l_z}{2} \Bigl ) + \delta\Bigr ( \Delta z + \dfrac{l_z}{2} \Bigl ) \right],
    \end{split}
\end{equation}
where $f\circledast g$ is the convolution of  functions $f$ and $g$, $\delta$ the Dirac delta function, and where for simplicity in writing $H_z$ we have taken the delta functions to be centered in $\pm l_z/2$ rather than in $\pm (l_z-l_\mathrm{surf}/2)/2$. Using the convolution theorem~\cite{goodman_speckle_2007-1}, $\mathcal{F}\left[G_z \circledast H_z\right] = \mathcal{F}\left[G_z\right] \mathcal{F}\left[H_z\right]$, and the identity $ \int_{-\infty}^{\infty}  \left[ \delta ( \Delta z - a ) + \delta( \Delta z + a ) \right] e^{iq \Delta z} \mathrm{d}\Delta z = 2\cos(q_z a) $, we obtain
\begin{equation}
    \begin{split}
        g_2(\vec{q}, \Delta\varepsilon) - 1 = 	\bigg | &\mathrm{sinc} \bigg(\frac{\Delta \varepsilon l_s q_x}{2} \bigg) \mathrm{sinc} \bigg(\frac{\Delta \varepsilon l_s  \nu q_y}{2} \bigg) \\ &\mathrm{sinc} \bigg(\frac{\Delta \varepsilon l_{\mathrm{surf}} \nu q_z}{2} \bigg)
        \mathrm{cos} \bigg(\frac{\Delta \varepsilon l_z \nu q_z}{2} \bigg)	 \bigg |^2 \,.
        \label{eq:surfacescatter}
    \end{split}
\end{equation}
Note that the last term in the r.h.s. of Eq.~\ref{eq:surfacescatter}, is responsible for the oscillations of $g_2(\vec{q},\Delta \epsilon)-1$, at a non-dimensional wavelength $\lambda_{\mathrm{osc}} = \dfrac{2 \pi}{q_z l_z \nu}$. 

The intensity correlation function shown in Fig.~\ref{fig:surface}a differs somehow from the form predicted by Eq.~\ref{eq:surfacescatter}, namely because the minima of the experimental $g_2-1$ never reach zero. This may be due to the oversimplified modeling adopted here: in particular, we neglected both bulk scattering and the attenuation of the incoming beam propagating through the sample. Nonetheless, to test whether Eq.~\ref{eq:surfacescatter} captures the essential features of the oscillations, we focus on their wavelength and compare the measured $\lambda_{\mathrm{osc}}$ to its theoretical value.

We perform experiments on samples of different thicknesses and plot the experimental wavelength of oscillations in the correlation function against $\dfrac{2 \pi}{q_z l_z \nu}$ in Fig.~\ref{fig:surface}b. The experimental $\lambda_{\mathrm{osc}}$ is in very good agreement with the theory, shown by the gray dashed line, with no fitting parameters. This confirms that surface scattering is responsible for the oscillations. Note that (very small) oscillations with the same wavelength are also seen in the FS experiment. Since $q_z$ for the FS configuration is much smaller than for the BS one, these oscillations can not originate from forward-scattered light. Rather, they are due to a small fraction of the incoming beam being reflected at the exit of the sample. This reflected beam counter-propagates in the sample and originates back-scattered light that is collected in the FS experiment, a phenomenon similar to that reported in Ref.~\cite{Pommella2019}.

To measure dynamics that come from the bulk of the sample and not from the surfaces, we make two changes to the apparatus. First, we greatly reduce the surface scattering, by immersing the sample in a pool filled with glycerol, whose refractive index is much closer to that of the sample as compared to air. The wedge shape of the pool, as seen in Fig.~\ref{fig:setup}a, is crucial to divert reflections of light scattered by the sample away from the cameras. 
Second, we increase scattering from the bulk by incorporating melamine formaldehyde colloidal particles, as detailed in Sec.~\ref{sec:materials}.
Repeating the experiments with these two changes results in correlation functions without oscillations for both BS and FS, as can be seen in Fig.~\ref{fig:surface}c. All data shown in the reminder of the paper were taken on particle-seeded PDMS samples immersed in glycerol.

We find that a few precautions should be taken in order to obtain well-reproducible correlation functions that decay in an approximately $\mathrm{sinc}^2(x)$-like way. First, measurements are best taken during a return travel after the sample has been pre-strained, i.e. for a set of negative $\Delta \epsilon$. This minimizes the risk of slip at the clamps: indeed, even a macroscopically undetectable slip can induce unpredictable variations in the correlation function, due to the sensitivity of the technique. Second, correlation functions are measured from a stationary region in the center of the sample as the two motors apply an equal (in magnitude) velocity to each end of the sample. This prevents a translation of the sample relative to glass wall of the pool, thus avoiding oscillations in $g_2-1$ that may arise from interferences between light scattered from the pool wall and from the sample. Indeed, the distance between the sample and the glass wall may vary during the test if both are not perfectly aligned to the direction of the applied strain, which experimentally is difficult to achieve.

\subsection{Tuning the decay rate of $g_2-1$ by varying the speckle size}
\label{sec:bulk}

The intensity correlation function for bulk scattering and purely affine motion is given by Eq.~\ref{eq:bulk_scatter}. An intriguing feature of this expressions is the dependence of the decay rate on the speckle size, owing to the $l_s$ term that appears in the first two $\mathrm{sinc}$ functions in the r.h.s., quantifying the lateral size of the scattering volume $V_s$. This is quite unusual: in conventional, far-field DLS, the detector is illuminated by light issued from the whole illuminated sample and the dynamics generally do not depend on the scattering volume, think, e.g. of Brownian motion. As a result, the decay rate of $g_2-1$ is independent of the collection optics and speckle size. Here, this effect is due to both the imaging configuration of PCI and the geometry of the mechanical test, with components of the strain gradient parallel to the imaged plane.

\begin{figure}[h]
    \centering
    \includegraphics[width=\linewidth]{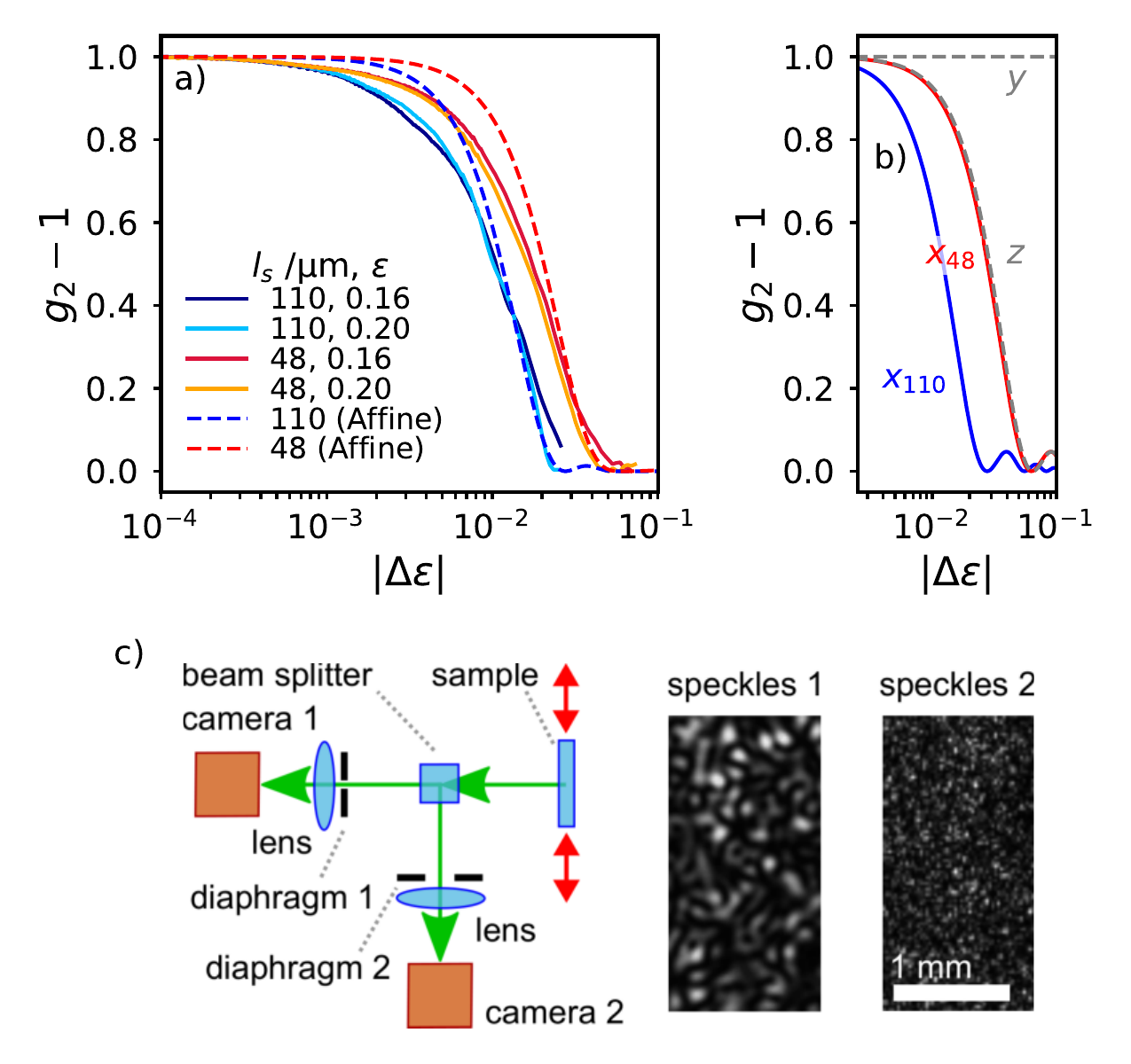}
    \caption{a) Correlation functions for two experiments (solid lines), with two different speckle sizes for a PDMS sample $1.77~\mathrm{mm}$ in thickness, measured  at the FS$_x$ scattering vector. The dashed lines are the theoretical prediction for purely affine motion, Eq.~\ref{eq:bulk_scatter}. b) Separate plot of each of the three sinc$^2$ factors of the theoretical $g_2-1$, for the experimental confditions of the data shown in a). Here, only the $x$ component varies with speckle size (blue and red lines) for the different speckle sizes. 
    c) The modified setup to collect the same scattered light with different speckle sizes.}
    \label{fig:speckle_size}
\end{figure}

To test experimentally the dependence of $g_2-1$ on $l_s$, we modify the experimental apparatus shown in Fig.~\ref{fig:setup}. As illustrated in Fig.~\ref{fig:speckle_size}c, we insert a beam splitter along the FS$_x$ collection path and move the FS$_y$ camera and its objective on one of the light paths exiting the beam splitter. With this arrangement, both cameras collect the same light scattered in the FS$_x$ direction, but the diaphragms placed in front of their respective objective lenses are set to a different aperture, so that the speckles for camera 1 are larger than those for camera 2, see the images in Fig.~\ref{fig:speckle_size}c. Figure~\ref{fig:speckle_size}a shows, for two independent experiments at different pre-strains $\varepsilon$, how the correlation functions measured at $\vec{q}(\mathrm{FS}_x)$ change with speckle size (solid lines). The magnitude of the change of the decay rate with speckle size agrees very well with theory, confirming the dependence on $l_s$. Interestingly, this indicates that one can tune the sensitivity of the setup to strain-induced microscopic dynamics simply by varying the speckle size, i.e. by opening or closing the objective diaphragms.

Figure~\ref{fig:speckle_size}a also shows that the experimental correlation functions generally lay below the correspondent theoretical curves (dashed lines), calculated using Eq.~\ref{eq:bulk_scatter}. This suggests that, at the $\qty{3}{\micro \meter}$ length scale probed by FS, there is a non-affine microscopic response to strain in the polymer network, in addition to affine displacements, agreeing with theory~\cite{Didonna2005,Heussinger2007} and experiments~\cite{Naumova2019, Basu2011, Wen2012, Hepworth2001} on a variety of polymer networks. Note that these non-affine dynamics are better seen in the initial part of $g_2-1$, corresponding to localized displacements, while the tail of the experimental correlation function is closer to the theoretical expression for affine motion. This suggests that non-affinity are more pronounced on smaller length scales, a feature that will be discussed in Sec.~\ref{sec:pre-strain}.

Some clues on the nature of these non-affine dynamics are provided by the $l_s$ dependence of the curves. Visual inspection of the data reveals that the decay of the measured correlation functions relative to their theoretical affine counterparts is similar for both speckle sizes. This rules out diffusive, random-walk-like non-affine dynamics. This is because the $g_2 - 1$ decay for random motion does not depend on speckle size. 
Therefore, if the non-affine
motion was purely diffusive, the relative decays of the measured
and affine $g_2 - 1$ would not be the same for two different speckle sizes. By contrast, the $l_s$ dependence of non-affinity suggest that these dynamics are more pronounced when probing a larger scattering volume, as for affine displacements, hinting at some degree of correlation between the magnitude of affine and non-affine displacements. This is similar to what found by optical microscopy in polyacrylamide gels, where the non-affine mean squared displacement of tracer particles was found to increase as the square of the applied strain~\cite{Basu2011}. 

As a final remark on the speckle size dependence of $g_2 - 1$, we note that this effect is most relevant when probing thin samples in forward scattering, i.e. for small $l_z$ and $q_z$. Indeed, a close look at Eq.~\ref{eq:bulk_scatter} shows that if $l_s q_x$ or $l_s \nu q_y$ are smaller than $l_z \nu q_z$, the decay of $g_2-1$ is ruled by the speckle-size independent $\mathrm{sinc} \left(\frac{\Delta \varepsilon l_z \nu q_z}{2} \right)$ term. To illustrate this point, we plot separately the components of the theoretical affine $g_2 - 1$ decay in Fig.~\ref{fig:speckle_size}b, for the conditions of our experiment (FS$_x$ scattering geometry and sample thickness $l_z=1.77~\mathrm{mm}$). One can see that for the larger speckle size, $l_s = \qty{110}{\micro\meter}$ (blue curve), the decay of the sinc$^2$ term containing $q_x$ (``$x$ term'') is faster than those due to the other terms, resulting in a marked dependence on $l_s$. As the speckle size decreases, the decay of the $x$ term becomes slower. For the smallest tested speckle size, $l_s = \qty{48}{\micro\meter}$, the decay rate of the $x$ and $z$ terms has become comparable. For even smaller speckles, we expect the overall decay of $g_2-1$ to be ruled by the $z$ term, which does not contain $l_s$, and hence to be independent of the speckle size.

\subsection{Effect of pre-strain on microscopic motion}
\label{sec:pre-strain}
\begin{figure}
    \centering
    \includegraphics[width=\linewidth]{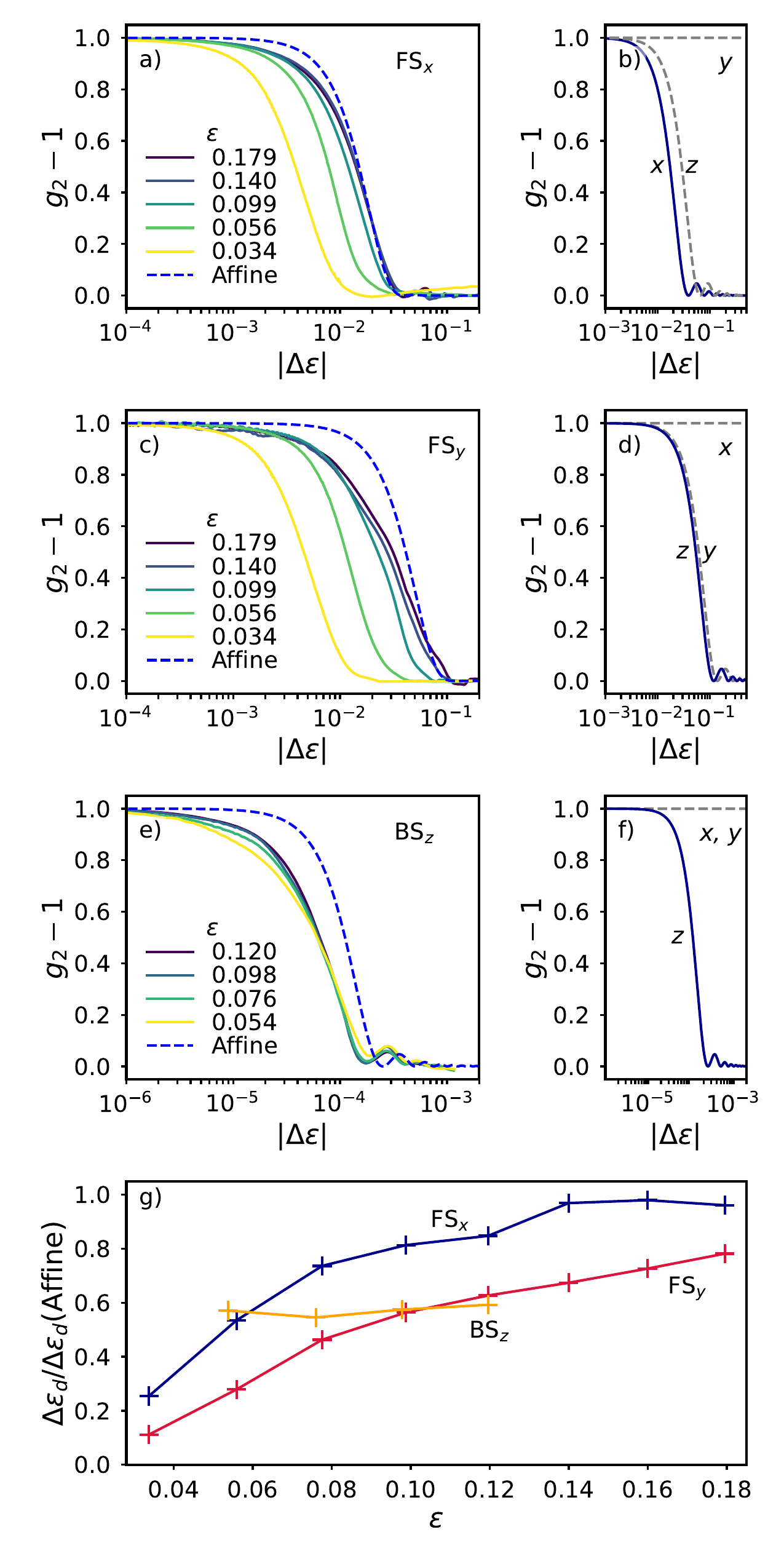}
    \caption{
    a), c), and e): correlation functions for several pre-strain values, for the scattering vectors FS$_x$, FS$_y$ and BS$_z$, respectively, for a $1.77~\mathrm{mm}$ thick PDMS sample. Data at the three scattering angles were collected simultaneously. Panels b), d) and f), correspondingly show the $x, y$ and $z$ sinc$^2$ terms of Eq.~\ref{eq:bulk_scatter} for FS$_x$, FS$_y$ and BS$_z$, respectively. The quickest decay dictates which component of motion is expected to dominate the decay of $g_2-1$. The full correlation functions for purely affine motion, i.e. the product of the three terms displayed in each panel b), d), f), are shown in a), c), e) as blue dashed lines. g) Ratio of the measured and theoretical characteristic decay strain increments, where $g_2-1$ drops to half of its initial value.}
    \label{fig:pre-strain}
\end{figure}

Our experiments revealed non-affine microscopic movements on a $\approx \qty{}{\micro \meter}$ length scale as a response to macroscopic strain in the bulk of our polymer sample. To better characterize this motion we repeated our experiments using the setup in the configuration of Fig. \ref{fig:setup} (using FS$_x$, FS$_y$  and BS$_z$ detectors and fixed speckle sizes of $76$, $56$ and $46~\mu$m, respectively) and varied the pre-strain $\epsilon$ starting from which the correlation functions are calculated. We start by examining $g_2-1$ for the FS$_x$ scattering vector, shown in Fig.~\ref{fig:pre-strain}a. In Fig.~\ref{fig:pre-strain}b, we plot separately the three factors of Eq.~\ref{eq:bulk_scatter} showing that, under purely affine motion, the decorrelation at this scattering vector is dominated by microscopic movements parallel to the direction of strain, $x$. Remarkably, the experimental correlation functions exhibit a dependence on pre-strain: at small $\varepsilon$, the experimental curves decay much faster than the theoretical prediction for affine dynamics (dashed blue line), while they get increasingly close to the form predicted by Eq.~\ref{eq:delta_xyz} as the pre-strain increases. 

We quantify the importance of non-affine motion as a function of pre-strain $\epsilon$ in Fig.~\ref{fig:pre-strain}g, where we plot $\widetilde{\Delta \epsilon_d}$, the characteristic strain increment over which the experimental $g_2-1$ functions decay, $\Delta \epsilon_d$, normalized by the corresponding $\Delta \epsilon_d$(Affine) for the affine theoretical expression, Eq.~\ref{eq:bulk_scatter}. $\Delta \epsilon_d$ is defined as the strain increment where $g_2-1=0.5$. For the FS$_x$ data, $\widetilde{\Delta \epsilon_d} = 0.27$ at low pre-strain, $0.034$ 
demonstrating large non-affine motion. However, at pre-strains above $\varepsilon > 0.14$ $\Delta \epsilon_d$ becomes $0.94$ of the purely affine value, showing a transition to mostly affine motion. The non-affinity seen here cannot be due to strain-induced damage, because in that case one would expect non-affine rearrangements to become increasingly important as the network is strained, contrary to the trend observed here. Furthermore, we recall that the range of $\epsilon$ probed here falls well within the mechanical linear regime, see Fig.~\ref{fig:setup}a. Rather, we propose that non-affine motion at low strain is due to the gradual unraveling of polymer segment coils between cross-links. As $\epsilon$ grows, the polymer chains become increasingly aligned along the stretching direction, loosing degrees of freedom associated with motion along $y$ and $z$, which results in microscopic dynamics close to purely affine. This effect tends to saturate for $\epsilon \geq 0.14$, consistent with our observations in Fig.~\ref{fig:speckle_size}a, where the dynamics measured at $\epsilon = 0.16$ and $0.20$ were essentially identical. Note that at these strains the polymer chains are certainly not completely uncoiled, otherwise one would observe strain hardening due to the transition from entropic to enthalpic elasticity, a feature not seen in Fig.~\ref{fig:setup}a).

We report in Fig.~\ref{fig:pre-strain}c) the correlation functions measured in the FS$_y$ geometry, for which the magnitude of the scattering vector is essentially the same as for $q(\mathrm{FS}_x)$, but its main component is oriented along $y$, perpendicular to the stretching direction. As a reference, we show in Fig.~\ref{fig:pre-strain}d the three terms of the purely affine $g_2-1$, whose decay is equally dominated by displacements parallel to the thickness of the sample, along $z$, and along $y$, since for FS$_y$ $q_yl_s \approx q_zl_z$. The overall trend of the experimental data is similar to that seen for FS$_x$: significant non-affinity is detected at small pre-strain, while at larger $\epsilon$ the data approach the theory for purely affine dynamics. Contrary to the motion along $x$, in the $y$ and $z$ direction polymer coils do not unravel with increasing strain. In fact, due to the positive Poisson's ratio, the polymer chains become more compressed along $y$ and $z$. However, again, the measured decays approach the affine decay (blue dashed line) as pre-strain is increased, indicating  a decrease in non-affine motion. We suggest that the combination of stretching along $x$ and compression along $z$ and $y$ increasingly confines the motion of polymer segments and results in less non affinity. The ratio of the measured and theoretical affine decay strains for FS$_y$ is shown in Fig.~\ref{fig:pre-strain}g (red line and symbols). We observe a similar trend but a curve translated downwards when compared to FS$_x$, indicating a higher non-affine contribution at all strains. This suggests a smaller loss of degree of freedoms of the polymer chains in the direction of compression, as compared to that in the stretching direction. 

Consistent with the data of Fig.~\ref{fig:speckle_size}a discussed above, we observe that for the FS correlation functions of Figs.~\ref{fig:pre-strain}a,b deviations with respect to purely affine dynamics are more pronounced at small strain increments, i.e. on small length scales. To directly asses the length scale dependence of non affinity, we plot in Fig.~\ref{fig:pre-strain}e the BS$_z$ experimental $g_2-1$ functions, and in Fig.~\ref{fig:pre-strain}
f the corresponding three terms of the theoretical purely affine dynamics. The length scale explored in the BS geometry is $\approx 90$nm, and the decay of $g_2-1$ is dominated by motion along $z$. In contrast to the FS data, in Fig.~\ref{fig:pre-strain}e we see that the correlation functions are far less dependent on pre-strain: at all the tested $\Delta \epsilon$, $g_2-1$ decays significantly faster than the affine prediction, with very little changes with $\epsilon$. The relative importance of non-affine dynamics is shown in Fig.~\ref{fig:pre-strain}g, confirming no significant dependence on strain, exhibiting a value of 0.57 across all measured $\epsilon$. Note that higher strains were inaccessible due to a high sensitivity to sample slip at the clamps in BS. Therefore, at a $90~\mathrm{nm}$ length scale, significant non-affine motion persists during stretching. This observation is consistent with the notion that non-affine motion is more prevalent at smaller length scales and agrees with findings for hydrogels under shear~\cite{Basu2011}. For our PDMS samples at a pre-strain $\epsilon \approx 0.14$, the FS and BS data collectively indicate that affine displacements are recovered on length scales of the order of a few microns. Thus, the cross-over between microscopic non-affinity and the macroscopic affine deformation expected for an elastic sample occurs on length scales surprisingly larger than the network mesh size. The dependence on $\epsilon$ uncovered here suggests that this cross-over may occur at even larger length scales upon decreasing the pre-strain.

\section{Discussion and conclusions}
\label{sec:conclusion}
We have described a new experimental apparatus to probe microscopic motion for uni-axially strained free-standing polymer network samples. The setup combines measurements of the macroscopic strain and stress with multispeckle, space and time-resolved dynamic light scattering. It is optimized for single scattering, which allows for resolving motion in specific directions, e.g. parallel or perpendicular to the stretching direction. However, we emphasize that the same optical layout could be used in the high multiple scattering regime of DWS, where the dual illumination scheme adopted here could be used for simultaneous DWS measurements in back-scattering and transmission. 

Working in the single scattering regime requires special care. In particular, we have shown that surface scattering may dominate the scattering signal for nearly transparent polymer samples. Unfortunately, in the configuration presented here surface scattering carries little useful physical information, besides the measurement of the Poisson's ratio along the sample thickness: the associated $g_2-1$ exhibits oscillations that only depend on the change of thickness during stretching. We have proposed and demonstrated an effective way to suppress the contribution of surface scattering, by immersing the sample in glycerol, to drastically reduce the refractive index jump at its surface, and by embedding tracer particles in the polymer network. At the low concentration of this work, tracer particles hardly affect the mechanical properties of the network and are slaved to it, thus faithfully reporting the polymer dynamics. At larger concentrations, filler particles are routinely used to improve the mechanical properties of, e.g., elastomers. Investigating their dynamics, especially in the non-linear regime, is a promising way to understand the relationship between microscopic rearrangements and macroscopic properties, e.g. in strain-induced aging~\cite{ehrburger-dolle_xpcs_2012}.

In experiments probing the dynamics under a time-evolving strain, it is important to quantify the contribution of affine displacements to the measured $g_2-1$, in order to discriminate between affine and non-affine dynamics, the latter typically being of greater interest. Unlike in small-angle scattering under shear, where it is possible to isolate non-affine motion by choosing $\vec{q}$ oriented in the vorticity direction~\cite{Aime2019}, for free-standing samples under tensile strain affine motion occurs in the three $x$, $y$, and $z$ directions. We have derived a simple expression for the affine $g_2-1$, containing sinc$^2$ terms similar to those already identified for far-field scattering under extension or shear, see e.g. Refs.~\cite{ehrburger-dolle_xpcs_2012,leheny_rheo-xpcs_2015,Aime2019}. However, the expression derived here, Eq.~\ref{eq:bulk_scatter}, differs from previous ones owing to the explicit dependence on two length scales: the sample thickness $l_z$ and the speckle size $l_s$. Because typically $l_z >> l_s$, the affine $g_2-1$ may be mostly sensitive to motion along the sample thickness, even for small angle scattering where the $z$ component of  $\vec{q}$ is much smaller than $q_x$ or $q_y$. A careful determination of the three components of the scattering vector and plots like those of Figs.~\ref{fig:speckle_size}b and~\ref{fig:pre-strain}b,d,f are useful ways to understand the relative importance of affine motion along the different directions for a given optical configuration. 

The dependence on speckle size $l_s$ is due to the peculiar imaging geometry of PCI. It can be leveraged to conveniently tune the sensitivity of the setup to microscopic motion: we have shown that closing the diaphragm of the PCI objective lens results in larger speckles and a faster decay of $g_2-1$ with $\Delta \epsilon$. Although not tested here, the setup could be easily modified to collect scattered light in the far field, by placing the cameras in the back focal plane of the objective lenses. In this case, $l_s$ would be replaced by the lateral size $l_{xy}$ of the illuminated sample volume in the $x,y$ plane, which would typically be comparable to or larger than the sample thickness. As a result, the FS $g_2-1$ would be sensitive essentially only to motion along $x$ or $y$, because one would have $q_x l_{xy} >> q_z l_z$ or $q_y l_{xy} >> q_z l_z$, for cameras 1 or 2, respectively. The far field geometry is also used in XPCS, where the beam size, however, is typically of the order of a few tens of microns at most. For XPCS, we thus expect the same kind of analysis of the relative weight of motion along $x$, $y$, and $z$ to be important. As a final remark on the contribution of affine displacements, we note that they are relevant only if $g_2-1$ is measured while the sample is macroscopically deformed, as in this work. For other protocols, such as stress relaxation at fixed imposed strain~\cite{ehrburger-dolle_xpcs_2012}, or the echo protocol where a sinusoidal deformation is imposed and the dynamics are probed stroboscopically at each cycle~\cite{hebraud_yielding_1997,hohler_periodic_1997,petekidis_rearrangements_2002}, the macroscopic strain between successive speckle images is zero and the affine deformation field vanishes. 

As an example of the possibilities of the new setup, we investigated the microscopic dynamics of PDMS networks. We unveiled substantial non-affine dynamics even at very low strains, well within the mechanical linear regime. By measuring the dynamics at different $q$ vectors, we have shown that the microscopic dynamics are strongly non-affine at short length scales $\approx 90$~nm. A cross-over to affine displacements is observed at larger length scales, of the order of a few microns. Our results show that non-affine displacements extend over length scales surprisingly larger than the mesh size, likely due to the disordered nature of the network. Remarkably, non-affine displacements decrease with increasing pre-strain, an effect that we attributed to polymer chains becoming less coiled-up as pre-strain increases. Finally, the speckle size dependence of $g_2-1$ has revealed that these non-affine dynamics are not fully decorrelated from affine displacements, in that the relative displacement between scatterers increase with the size of the probed sample volume, as for affine motion.

We end by noting that this apparatus may be applied to any free-standing and (nearly) transparent material, such as, e.g., biological samples, elastomers and hydrogels, to examine their microscopic responses under tensile tests. While in this work we demonstrated the new setup in the rheological linear regime, we expect its capabilities to be fully exploited in the non-linear regime, since localized damage may be detected at a microscopic scale thanks to the space-resolved features of PCI. We are currently leveraging these capabilities to investigate the spatio-temporal evolution of the microscopic dynamics at the onset of crack propagation in pre-notched multiple elastomer networks~\cite{Ducrot2014}.

\section*{Author contributions}
N. H. P. Orr: Conceptualization, methodology, software, investigation, formal analysis, writing - original draft. G. Prevot: Resources, software. T. Phou: Resources, sample synthesis. L. Cipelletti: Conceptualization, methodology, software, investigation, writing - review $\&$ editing, supervision, funding acquisition.

\section*{Conflicts of interest}
There are no conflicts to declare.

\section*{Data availability}
The data is available on the Zenodo repository (doi: 10.5281/zenodo.16811288.)

\section*{Acknowledgements}
We thank J. Barbat for help in instrumentation, and L. Ramos, C. Creton, J. L. Barrat, and K. Martens for discussions. This work was supported by ANR, grant MultiNet ANR-20-CE06-0028. LC gratefully acknowledges support by the Institut Universitaire de France.

\section{Supplementary Information}
\subsection{Impact of the addition of melamine resin microparticles on the mechanical properties of PDMS samples}
The stress response to uni-axial strain of identically prepared samples but without the addition of melamine resin microparticles is shown in Fig.~ SI1. We find a difference of $12 \%$ between the Young's moduli in similar samples with and without microparticles. Here, microparticle addition lowered the Young's modulus in the samples we considered, where often microparticle addition has the opposite effect. This may indicate that particles have no significant effect, the difference seen here rather resulting from slight sample-to-sample variations of the Young's modulus due to the synthesis protocol.

\begin{figure}[h]
    \centering
    \includegraphics[width=\linewidth]{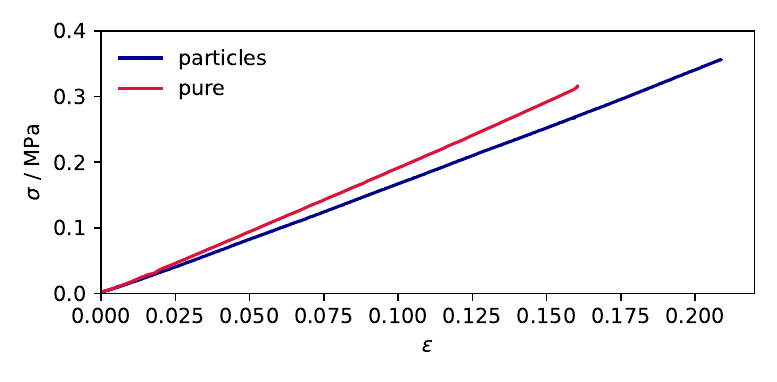}
    \caption*{Fig. SI1. True stress against true strain for PDMS samples prepared with microparticles, blue, and without, red. }
    \label{fig:pureparticles}
\end{figure}

\subsection{Purely affine microscopic displacement under uni-axial strain}

We consider the microscopic displacement of a scatterer between times $t$ and $t+\Delta t$ for a sample undergoing a purely affine uni-axial extension or contraction along the $x$ direction. We start by considering motion along $x$. Let $x(t)$ be the $x$ coordinate of the scatterer in a reference frame where the origin is fixed (e.g., $O$ coincides with the center of a sample equally stretched from both sides). For a purely affine displacement, one has 
\begin{equation}
x(t+\Delta t) - x(t) = x(t) \frac{L(t+\Delta t)}{L(t)} - x(t) = x(t) \frac{\Delta L(t,\Delta t)}{L(t)} \,,
\label{eq:x_displacement}
\end{equation}
with $\Delta L(t,\Delta t)=L(t+\Delta t)-L(t)$ and $L$ the sample dimension along $x$. The fraction in the last term on the r.h.s. of Eq.~\ref{eq:x_displacement} is the incremental engineering strain, i.e. the strain over $\Delta t$ calculated taking as a reference $L(t)$, the length at time $t$, not the initial length, $L_0$. It is easy to show that this quantity is equal, to leading order in $\Delta L /L$, to the increment of the true strain, $\Delta \varepsilon(t,\Delta t) = \varepsilon(t+\Delta t) - \varepsilon(t)$:
\begin{equation}
\Delta \varepsilon = \ln{\frac{L(t)+\Delta L}{L_0}} - \ln{\frac{L(t)}{L_0}} = \ln{\frac{L(t)+\Delta L}{L(t)}} = \frac{\Delta L}{L(t)} + \mathcal{O}\left[\frac{\Delta L}{L(t)}\right]^2\,.
\label{eq:true_strain_increment}
\end{equation}
In this paper, we consistently use true strain to quantify deformation. As shown in the main text, $g_2-1$ typically fully decays over true strain increments $\Delta \varepsilon << 1$. Hence, one can safely neglect the difference between incremental engineering strain and $\Delta \varepsilon$. Accordingly, using Eqs.~\ref{eq:x_displacement} and ~\ref{eq:true_strain_increment}, we express the scatterer displacement as
\begin{equation}
    x(t+\Delta t)-x(t) \equiv \Delta x =  x(t)\Delta \varepsilon(t,\Delta t) \,.
\end{equation}
The same arguments apply to displacements along the $y$ and $z$ directions, provided that the Poisson's ratio $\nu$ is used to account for the difference in response to stretching for the different directions. One thus finds
\begin{equation}
    \begin{split}
        \Delta x &= x\Delta\varepsilon \\
        \Delta y &=  -\nu y \Delta\varepsilon \\
        \Delta z &=  -\nu z \Delta\varepsilon  \,,
    \end{split}
\label{eq:SIdelta_xyz}
\end{equation}
which are Eqs. 3 of the main text.


\balance




\providecommand*{\mcitethebibliography}{\thebibliography}
\csname @ifundefined\endcsname{endmcitethebibliography}
{\let\endmcitethebibliography\endthebibliography}{}

\end{document}